%% file: gsh-levy.tex
\documentstyle[twoside,fleqn,espcrc2]{article}

\newcommand{\AmS}{{\protect\the\textfont2
  A\kern-.1667em\lower.5ex\hbox{M}\kern-.125emS}}

\title{Do we observe L\'evy flights in cosmic rays?}

\author{G. Wilk\address{The Andrzej Soltan Institute for Nuclear Studies, 
        Nuclear Theory Department, Warsaw, Poland\\
        email: wilk@fuw.edu.pl}%
	\thanks{Talk given  at the Xth International Symposium
	on Very High Energy Cosmic Ray Interactions, Laboratori
	Nazionali del Gran Sasso, 12-17 July 1998, Assergi,
	Italy, to be published in the proceedings ({\sl Nucl.
	Phys. {\bf B} (Proc. Suppl.)).}}
        and 
        Z.W\l odarczyk\address{Institute of Physics, Pedagogical 
        University, Kielce, Poland \\
        email: wlod@pu.kielce.pl}}
       
\begin{document}

\begin{abstract}
We argue that the so called {\it long flying component} (LFC)
observed in some cosmic ray experiments are yet another manifestation
of L\'evy distributions (with index $q=1.3$), this time of the
distribution observation probability of the depths of starting points
of cascades. It means that LFC is governed by the so called long-tail
L\'evy-like anomalous superdiffusion, a phenomenon frequently
encountered in Nature. Its connection with the so called 
Tsallis's statistics is also briefly discussed. 
\end{abstract}

\maketitle

\section{INTRODUCTION}

Cosmic ray experiments report for some time the existence of the
apparent {\it long flying component} (LFC) phenomenon in the
propagation of the initial flux of incoming nucleons. We shall
concentrate here on the example of the distribution of cascade
starting points in the extra thick lead chamber of the Pamir
experiment (cf. \cite{WW} for detailed references concerning the
whole subject considered here). What is regarded as peculiar or even
unexpected is the fact that instead of the normaly expected
exponential fall off of the depth distribution of the starting points
of cascades  

\begin{equation}
\frac{d\, N(T)}{d\, T}\, =\, {\rm const}\cdot \exp\left( -
                T/\lambda\right)  \label{eq:EXP}
\end{equation}

\noindent
one gets power law behaviour of the type (cf. Fig. 1):

\begin{equation}
\frac{d\, N(T)}{d\, T}\, =\, {\rm const} \cdot
    \left [ 1 - \frac{(1 - q)\, T}{\lambda}\right]^{\frac{1}{1-q}}
\label{eq:POWER}
\end{equation}

\noindent
with parameter $q$ (to be specified below) equal $q=1.3$ and with the
absorption length (in $c.u.$) for the interaction of hadrons in the
lead chamber equal to  $\lambda = 18.85\pm 6.62$). The
immediate question one is faced with is: {\it have somebody seen
something like this somewhere?}. The answer to it is positive and we
shall concentrate on it in what follows.

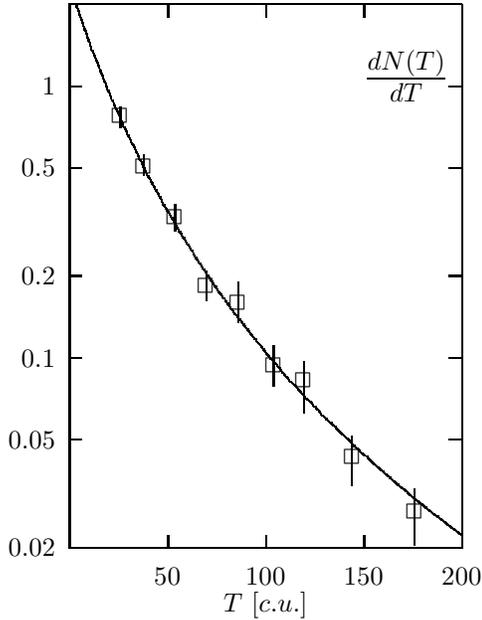
\begin{figure}[hbt]
\vspace{9pt}
\input{fig_levy}
\caption{Depth distribution of the starting points of cascades in 
Pamir lead chamber. Notice the non-exponential behaviour of
data points (for their origin cf. [1]).}
\end{figure}

\section{L\'EVY DISTRIBUTIONS}

In many places (albeit in different circumstances than encountered
here like, for example, distributions of heartbeats, travel patterns
of albatroses, behaviour of stock market, time dependence of the
leaky faucet and numerous other) one discoveres that the respective
distributions are of the type

\begin{equation}
P(x) \sim \frac{1}{x^{1 + \gamma}} \qquad {\rm for~large}~x .
\label{eq:L} 
\end{equation}

\noindent
They represent the so called L\'evy distributions $L_{\gamma}(x)$ the
characteristic feature of which is that they allow for much longer
jumps (called {\it L\'evy flights}) than in normal diffusion
\cite{Ts}. Whereas in normal diffusion emerging from the Brownian
motion one encounters chaotic (usually small) jumps governed by
Gaussian distribution with finite variance, in L\'evy flights long
jumps appear intermittently with short ones and variance is
divergent. As a result one gets a fractal structure of the
distribution of points visited by a random walker as demonstrated in
\cite{F}. 

\section{SIMPLE-MINDED DERIVATION}

The only possible way to (formaly) reconcile the power-like behaviour
of eq.(\ref{eq:POWER}) with the exponential eq.(\ref{eq:EXP}) is to
make substitution $\lambda \rightarrow \lambda(T) = \lambda - (1-q)T
= \lambda + 0.3 T$ in the later. Notice that our previous explanation
of the LFC \cite{WW} which was based on the idea of fluctuating cross
sections leads {\it effectively} to such a replacement. Performing
now such replacement in the derivation of eq.{\ref{eq:EXP}) one gets

\begin{equation}
\frac{dN}{N}\, =\, - \frac{dx}{\lambda}\quad \Rightarrow \quad
                   - \frac{dx}{\lambda \left[1 - (1 -
                   q)\frac{x}{\lambda}\right]} . \label{eq:Nx}
\end{equation}                   

\noindent
Changing now variables to $z = 1 - (1-q)\frac{x}{\lambda}$ one gets

\begin{equation}
\frac{dN}{N}\, =\, \frac{1}{1-q}\, \frac{dz}{z}\quad \Rightarrow \quad
           N\, =\, N_0\, z^{\frac{1}{1-q}}, \label{eq:Nz}
\end{equation}

\noindent
i.e., equation (\ref{eq:POWER}).           

\section{TSALLIS'S STATISTICS}

The whole above discussion can be put on more formal level by
introducing the concept of information entropy

\begin{equation}
S\, =\, -\, \sum_i\, p_i\, \ln p_i \label{eq:IE}
\end{equation}

\noindent
for a given probability distribution $p_i$. It can be shown (cf.
\cite{INWW} for relevant references and discussion for production
processes) that the most plausible or least biased model independent
$p_i$'s for a given process can be obtain by maximalizing $S$ under
constraints: 

\begin{equation}
\sum p_i\, =\, 1\qquad {\rm and}\quad 
\sum R^{(l)}_i p_i\, =\, <\hat{R}^{(l)}>, \label{eq:constr}
\end{equation}

\noindent
where first one assures proper normalization of the probability
distribution $p_i$ whereas the rest $l=1,2,\dots $ of them represent
our experimental knowledge on the problem under consideration
\cite{INWW}. In this way $p_i$ tells us {\it the truth, the whole
truth} about our experiment, i.e., reproduces known information but
also it tells us {\it nothing but the truth}, i.e., it conveys the
least information. The Gaussian diffusion can be then derived
straightforwardly \cite{L} by demanding finitness of the second moment
of $p(x)$. Hower, this formalism fails to describe random walks with
more complex jump probabilities (like L\'evy distributions). In
general it fails for all physical systems involving either long range
correlations or long-range memory effects or fractal space-time
structure. To overcome this problem Tsallis proposed generalised
statistics leading to generalised entropy \cite{Ts,TE}:

\begin{equation}
S_q\, =\, -k \frac{1\, -\, \sum p_i^q}{1\, -\, q} \label{eq:TE}
\end{equation}

\noindent
which in the limit when new parametr $q$ approaches unity,
$q\rightarrow 1$, goes into (\ref{eq:IE}) ($k$ is constant, which in
eq.(\ref{eq:IE}) was put equal unity). This entropy can be now used in
the same way as mentioned above with the $l=1,2,\dots$
constraints representing
measured observables changed to

\begin{equation}
\sum R^{q(l)}_i p^q_i \, =\, <\hat{R}_q^{(l)}>. \label{eq:newc}
\end{equation}

\noindent
In this manner we get instead the Gaussian distribution, as
before, the following power-like distribution \cite{TE}

\begin{equation}
p_q(x)\, =\, \frac{1}{Z_q}\left[ 1\, -\, \beta (1 - q)
                           x^2\right]^{\frac{1}{1-q}}  \label{eq:PQ}\\
\end{equation}

\noindent
where

\begin{equation}
Z_q = \int dx \left[1 - \beta(1-q)x^2\right]^{1/(1-q)} . 
\end{equation}

\noindent
Notice that for large $x$ distribution (\ref{eq:PQ}) 
with $q=(3+\gamma)/(1+\gamma)$ coincides with distribution given by
eq. (\ref{eq:L}) representing L\'evy flights. 

The new parametr $q$ introduced in (\ref{eq:TE}) describes in a
simple way possible correlations present in the system in the sense
that, as is obvious from the new form of the constraint equations, it
either enhances frequest events (for $q>1$) or enhances
the rare ones (for $q<1$). The best way to demonstrate this property
is to look at the entropy of the system composed with two 
subsystems $A$ and $B$:

\begin{eqnarray}
S_q(A\cup B) &=& S_q(A) + S_q(B) + \nonumber \\
             && \, +\,  (1-q) S_q(A)\cdot S_q(B) . \label{eq:SS}
\end{eqnarray}

\noindent
It is in obvious way non-additive now, the normal additivity is 
recovered only in the usual case of $q=1$ (leading to eq. 
(\ref{eq:IE})). It means that we are facing a nonextensitivity 
here: subadditivity for $q>1$ and superadditivity for $q<1$ 
\cite{TE,PP,Tnon}.

\section{SUMMARY AND CONCLUSIONS}

We have proposed a novel approach to the (regarded as "strange" 
or "suspicious") power-like behaviour of some observables measured
in cosmic ray experiments. The novely of this approach 
is in the fact that even without detail dynamical knowledge of 
the origin of such effects (as provided, for example, in \cite{WW})
it is still worth to parametrize them with a single new
parameter $q$. This parameter expresses summarily some
new, so far undiscovered, dynamics by showing that there
are some correlations present in these particular measurements. 
The final decifering of its meaning is out of the scope of 
this presentation and will be attempted elsewhere. 
Here we would only like to bring ones attention to the fact 
that in essentially the same way one can generalize the 
usual information entropy approach used to describe
single particle distributions in multiparticle production
processes (as provided for example, in \cite{INWW}) to the 
case of more general Tsalis's entropy $S_q$. The ultimate goal
in this case would be the unique description of leading particle 
spectra in terms of a single physical parameter $q$. 
This would be especially suited for description of a great amount
of cosmic ray data. Such project is at present under investigation.
Finally, let us mention that Tsallis's entropy can still
be generalised even more (cf. \cite{A}) but physical meaning 
of this generalization is, at least from the point of view 
of cosmic ray applications, not yet clear.

\end{document}

%% file: fig_levy.tex
\setlength{\unitlength}{0.240900pt}
\ifx\plotpoint\undefined\newsavebox{\plotpoint}\fi
\sbox{\plotpoint}{\rule[-0.200pt]{0.400pt}{0.400pt}}%
\begin{picture}(900,990)(0,0)
\font\gnuplot=cmr10 at 10pt
\gnuplot
\sbox{\plotpoint}{\rule[-0.200pt]{0.400pt}{0.400pt}}%
\put(220.0,113.0){\rule[-0.200pt]{0.400pt}{205.729pt}}
\put(220.0,113.0){\rule[-0.200pt]{4.818pt}{0.400pt}}
\put(198,113){\makebox(0,0)[r]{0.02}}
\put(816.0,113.0){\rule[-0.200pt]{4.818pt}{0.400pt}}
\put(220.0,283.0){\rule[-0.200pt]{4.818pt}{0.400pt}}
\put(198,283){\makebox(0,0)[r]{0.05}}
\put(816.0,283.0){\rule[-0.200pt]{4.818pt}{0.400pt}}
\put(220.0,411.0){\rule[-0.200pt]{4.818pt}{0.400pt}}
\put(198,411){\makebox(0,0)[r]{0.1}}
\put(816.0,411.0){\rule[-0.200pt]{4.818pt}{0.400pt}}
\put(220.0,540.0){\rule[-0.200pt]{4.818pt}{0.400pt}}
\put(198,540){\makebox(0,0)[r]{0.2}}
\put(816.0,540.0){\rule[-0.200pt]{4.818pt}{0.400pt}}
\put(220.0,710.0){\rule[-0.200pt]{4.818pt}{0.400pt}}
\put(198,710){\makebox(0,0)[r]{0.5}}
\put(816.0,710.0){\rule[-0.200pt]{4.818pt}{0.400pt}}
\put(220.0,838.0){\rule[-0.200pt]{4.818pt}{0.400pt}}
\put(198,838){\makebox(0,0)[r]{1}}
\put(816.0,838.0){\rule[-0.200pt]{4.818pt}{0.400pt}}
\put(374.0,113.0){\rule[-0.200pt]{0.400pt}{4.818pt}}
\put(374,68){\makebox(0,0){50}}
\put(374.0,947.0){\rule[-0.200pt]{0.400pt}{4.818pt}}
\put(528.0,113.0){\rule[-0.200pt]{0.400pt}{4.818pt}}
\put(528,68){\makebox(0,0){100}}
\put(528.0,947.0){\rule[-0.200pt]{0.400pt}{4.818pt}}
\put(682.0,113.0){\rule[-0.200pt]{0.400pt}{4.818pt}}
\put(682,68){\makebox(0,0){150}}
\put(682.0,947.0){\rule[-0.200pt]{0.400pt}{4.818pt}}
\put(836.0,113.0){\rule[-0.200pt]{0.400pt}{4.818pt}}
\put(836,68){\makebox(0,0){200}}
\put(836.0,947.0){\rule[-0.200pt]{0.400pt}{4.818pt}}
\put(220.0,113.0){\rule[-0.200pt]{148.394pt}{0.400pt}}
\put(836.0,113.0){\rule[-0.200pt]{0.400pt}{205.729pt}}
\put(220.0,967.0){\rule[-0.200pt]{148.394pt}{0.400pt}}
\put(749,855){\makebox(0,0){\Large{$\frac{dN(T)}{dT}$}}}
\put(528,23){\makebox(0,0){$T~[c.u.]$}}
\put(220.0,113.0){\rule[-0.200pt]{0.400pt}{205.729pt}}
\put(230.67,963){\rule{0.400pt}{0.964pt}}
\multiput(230.17,965.00)(1.000,-2.000){2}{\rule{0.400pt}{0.482pt}}
\multiput(232.59,958.32)(0.485,-1.332){11}{\rule{0.117pt}{1.129pt}}
\multiput(231.17,960.66)(7.000,-15.658){2}{\rule{0.400pt}{0.564pt}}
\multiput(239.59,939.60)(0.482,-1.575){9}{\rule{0.116pt}{1.300pt}}
\multiput(238.17,942.30)(6.000,-15.302){2}{\rule{0.400pt}{0.650pt}}
\multiput(245.59,921.88)(0.482,-1.485){9}{\rule{0.116pt}{1.233pt}}
\multiput(244.17,924.44)(6.000,-14.440){2}{\rule{0.400pt}{0.617pt}}
\multiput(251.59,904.88)(0.482,-1.485){9}{\rule{0.116pt}{1.233pt}}
\multiput(250.17,907.44)(6.000,-14.440){2}{\rule{0.400pt}{0.617pt}}
\multiput(257.59,888.79)(0.485,-1.179){11}{\rule{0.117pt}{1.014pt}}
\multiput(256.17,890.89)(7.000,-13.895){2}{\rule{0.400pt}{0.507pt}}
\multiput(264.59,872.16)(0.482,-1.395){9}{\rule{0.116pt}{1.167pt}}
\multiput(263.17,874.58)(6.000,-13.579){2}{\rule{0.400pt}{0.583pt}}
\multiput(270.59,856.16)(0.482,-1.395){9}{\rule{0.116pt}{1.167pt}}
\multiput(269.17,858.58)(6.000,-13.579){2}{\rule{0.400pt}{0.583pt}}
\multiput(276.59,840.43)(0.482,-1.304){9}{\rule{0.116pt}{1.100pt}}
\multiput(275.17,842.72)(6.000,-12.717){2}{\rule{0.400pt}{0.550pt}}
\multiput(282.59,825.43)(0.482,-1.304){9}{\rule{0.116pt}{1.100pt}}
\multiput(281.17,827.72)(6.000,-12.717){2}{\rule{0.400pt}{0.550pt}}
\multiput(288.59,811.26)(0.485,-1.026){11}{\rule{0.117pt}{0.900pt}}
\multiput(287.17,813.13)(7.000,-12.132){2}{\rule{0.400pt}{0.450pt}}
\multiput(295.59,796.71)(0.482,-1.214){9}{\rule{0.116pt}{1.033pt}}
\multiput(294.17,798.86)(6.000,-11.855){2}{\rule{0.400pt}{0.517pt}}
\multiput(301.59,782.71)(0.482,-1.214){9}{\rule{0.116pt}{1.033pt}}
\multiput(300.17,784.86)(6.000,-11.855){2}{\rule{0.400pt}{0.517pt}}
\multiput(307.59,768.99)(0.482,-1.123){9}{\rule{0.116pt}{0.967pt}}
\multiput(306.17,770.99)(6.000,-10.994){2}{\rule{0.400pt}{0.483pt}}
\multiput(313.59,756.50)(0.485,-0.950){11}{\rule{0.117pt}{0.843pt}}
\multiput(312.17,758.25)(7.000,-11.251){2}{\rule{0.400pt}{0.421pt}}
\multiput(320.59,742.99)(0.482,-1.123){9}{\rule{0.116pt}{0.967pt}}
\multiput(319.17,744.99)(6.000,-10.994){2}{\rule{0.400pt}{0.483pt}}
\multiput(326.59,730.26)(0.482,-1.033){9}{\rule{0.116pt}{0.900pt}}
\multiput(325.17,732.13)(6.000,-10.132){2}{\rule{0.400pt}{0.450pt}}
\multiput(332.59,717.99)(0.482,-1.123){9}{\rule{0.116pt}{0.967pt}}
\multiput(331.17,719.99)(6.000,-10.994){2}{\rule{0.400pt}{0.483pt}}
\multiput(338.59,705.26)(0.482,-1.033){9}{\rule{0.116pt}{0.900pt}}
\multiput(337.17,707.13)(6.000,-10.132){2}{\rule{0.400pt}{0.450pt}}
\multiput(344.59,693.74)(0.485,-0.874){11}{\rule{0.117pt}{0.786pt}}
\multiput(343.17,695.37)(7.000,-10.369){2}{\rule{0.400pt}{0.393pt}}
\multiput(351.59,681.54)(0.482,-0.943){9}{\rule{0.116pt}{0.833pt}}
\multiput(350.17,683.27)(6.000,-9.270){2}{\rule{0.400pt}{0.417pt}}
\multiput(357.59,670.54)(0.482,-0.943){9}{\rule{0.116pt}{0.833pt}}
\multiput(356.17,672.27)(6.000,-9.270){2}{\rule{0.400pt}{0.417pt}}
\multiput(363.59,659.26)(0.482,-1.033){9}{\rule{0.116pt}{0.900pt}}
\multiput(362.17,661.13)(6.000,-10.132){2}{\rule{0.400pt}{0.450pt}}
\multiput(369.59,647.98)(0.485,-0.798){11}{\rule{0.117pt}{0.729pt}}
\multiput(368.17,649.49)(7.000,-9.488){2}{\rule{0.400pt}{0.364pt}}
\multiput(376.59,636.82)(0.482,-0.852){9}{\rule{0.116pt}{0.767pt}}
\multiput(375.17,638.41)(6.000,-8.409){2}{\rule{0.400pt}{0.383pt}}
\multiput(382.59,626.54)(0.482,-0.943){9}{\rule{0.116pt}{0.833pt}}
\multiput(381.17,628.27)(6.000,-9.270){2}{\rule{0.400pt}{0.417pt}}
\multiput(388.59,615.82)(0.482,-0.852){9}{\rule{0.116pt}{0.767pt}}
\multiput(387.17,617.41)(6.000,-8.409){2}{\rule{0.400pt}{0.383pt}}
\multiput(394.59,605.82)(0.482,-0.852){9}{\rule{0.116pt}{0.767pt}}
\multiput(393.17,607.41)(6.000,-8.409){2}{\rule{0.400pt}{0.383pt}}
\multiput(400.59,596.21)(0.485,-0.721){11}{\rule{0.117pt}{0.671pt}}
\multiput(399.17,597.61)(7.000,-8.606){2}{\rule{0.400pt}{0.336pt}}
\multiput(407.59,585.82)(0.482,-0.852){9}{\rule{0.116pt}{0.767pt}}
\multiput(406.17,587.41)(6.000,-8.409){2}{\rule{0.400pt}{0.383pt}}
\multiput(413.59,575.82)(0.482,-0.852){9}{\rule{0.116pt}{0.767pt}}
\multiput(412.17,577.41)(6.000,-8.409){2}{\rule{0.400pt}{0.383pt}}
\multiput(419.59,566.09)(0.482,-0.762){9}{\rule{0.116pt}{0.700pt}}
\multiput(418.17,567.55)(6.000,-7.547){2}{\rule{0.400pt}{0.350pt}}
\multiput(425.59,557.21)(0.485,-0.721){11}{\rule{0.117pt}{0.671pt}}
\multiput(424.17,558.61)(7.000,-8.606){2}{\rule{0.400pt}{0.336pt}}
\multiput(432.59,547.09)(0.482,-0.762){9}{\rule{0.116pt}{0.700pt}}
\multiput(431.17,548.55)(6.000,-7.547){2}{\rule{0.400pt}{0.350pt}}
\multiput(438.59,538.09)(0.482,-0.762){9}{\rule{0.116pt}{0.700pt}}
\multiput(437.17,539.55)(6.000,-7.547){2}{\rule{0.400pt}{0.350pt}}
\multiput(444.59,529.09)(0.482,-0.762){9}{\rule{0.116pt}{0.700pt}}
\multiput(443.17,530.55)(6.000,-7.547){2}{\rule{0.400pt}{0.350pt}}
\multiput(450.59,520.09)(0.482,-0.762){9}{\rule{0.116pt}{0.700pt}}
\multiput(449.17,521.55)(6.000,-7.547){2}{\rule{0.400pt}{0.350pt}}
\multiput(456.59,511.45)(0.485,-0.645){11}{\rule{0.117pt}{0.614pt}}
\multiput(455.17,512.73)(7.000,-7.725){2}{\rule{0.400pt}{0.307pt}}
\multiput(463.59,502.37)(0.482,-0.671){9}{\rule{0.116pt}{0.633pt}}
\multiput(462.17,503.69)(6.000,-6.685){2}{\rule{0.400pt}{0.317pt}}
\multiput(469.59,494.09)(0.482,-0.762){9}{\rule{0.116pt}{0.700pt}}
\multiput(468.17,495.55)(6.000,-7.547){2}{\rule{0.400pt}{0.350pt}}
\multiput(475.59,485.37)(0.482,-0.671){9}{\rule{0.116pt}{0.633pt}}
\multiput(474.17,486.69)(6.000,-6.685){2}{\rule{0.400pt}{0.317pt}}
\multiput(481.59,477.69)(0.485,-0.569){11}{\rule{0.117pt}{0.557pt}}
\multiput(480.17,478.84)(7.000,-6.844){2}{\rule{0.400pt}{0.279pt}}
\multiput(488.59,469.37)(0.482,-0.671){9}{\rule{0.116pt}{0.633pt}}
\multiput(487.17,470.69)(6.000,-6.685){2}{\rule{0.400pt}{0.317pt}}
\multiput(494.59,461.37)(0.482,-0.671){9}{\rule{0.116pt}{0.633pt}}
\multiput(493.17,462.69)(6.000,-6.685){2}{\rule{0.400pt}{0.317pt}}
\multiput(500.59,453.37)(0.482,-0.671){9}{\rule{0.116pt}{0.633pt}}
\multiput(499.17,454.69)(6.000,-6.685){2}{\rule{0.400pt}{0.317pt}}
\multiput(506.59,445.37)(0.482,-0.671){9}{\rule{0.116pt}{0.633pt}}
\multiput(505.17,446.69)(6.000,-6.685){2}{\rule{0.400pt}{0.317pt}}
\multiput(512.00,438.93)(0.492,-0.485){11}{\rule{0.500pt}{0.117pt}}
\multiput(512.00,439.17)(5.962,-7.000){2}{\rule{0.250pt}{0.400pt}}
\multiput(519.59,430.37)(0.482,-0.671){9}{\rule{0.116pt}{0.633pt}}
\multiput(518.17,431.69)(6.000,-6.685){2}{\rule{0.400pt}{0.317pt}}
\multiput(525.59,422.37)(0.482,-0.671){9}{\rule{0.116pt}{0.633pt}}
\multiput(524.17,423.69)(6.000,-6.685){2}{\rule{0.400pt}{0.317pt}}
\multiput(531.59,414.65)(0.482,-0.581){9}{\rule{0.116pt}{0.567pt}}
\multiput(530.17,415.82)(6.000,-5.824){2}{\rule{0.400pt}{0.283pt}}
\multiput(537.00,408.93)(0.492,-0.485){11}{\rule{0.500pt}{0.117pt}}
\multiput(537.00,409.17)(5.962,-7.000){2}{\rule{0.250pt}{0.400pt}}
\multiput(544.59,400.65)(0.482,-0.581){9}{\rule{0.116pt}{0.567pt}}
\multiput(543.17,401.82)(6.000,-5.824){2}{\rule{0.400pt}{0.283pt}}
\multiput(550.59,393.37)(0.482,-0.671){9}{\rule{0.116pt}{0.633pt}}
\multiput(549.17,394.69)(6.000,-6.685){2}{\rule{0.400pt}{0.317pt}}
\multiput(556.59,385.65)(0.482,-0.581){9}{\rule{0.116pt}{0.567pt}}
\multiput(555.17,386.82)(6.000,-5.824){2}{\rule{0.400pt}{0.283pt}}
\multiput(562.59,378.65)(0.482,-0.581){9}{\rule{0.116pt}{0.567pt}}
\multiput(561.17,379.82)(6.000,-5.824){2}{\rule{0.400pt}{0.283pt}}
\multiput(568.00,372.93)(0.581,-0.482){9}{\rule{0.567pt}{0.116pt}}
\multiput(568.00,373.17)(5.824,-6.000){2}{\rule{0.283pt}{0.400pt}}
\multiput(575.59,365.65)(0.482,-0.581){9}{\rule{0.116pt}{0.567pt}}
\multiput(574.17,366.82)(6.000,-5.824){2}{\rule{0.400pt}{0.283pt}}
\multiput(581.59,358.65)(0.482,-0.581){9}{\rule{0.116pt}{0.567pt}}
\multiput(580.17,359.82)(6.000,-5.824){2}{\rule{0.400pt}{0.283pt}}
\multiput(587.59,351.65)(0.482,-0.581){9}{\rule{0.116pt}{0.567pt}}
\multiput(586.17,352.82)(6.000,-5.824){2}{\rule{0.400pt}{0.283pt}}
\multiput(593.00,345.93)(0.581,-0.482){9}{\rule{0.567pt}{0.116pt}}
\multiput(593.00,346.17)(5.824,-6.000){2}{\rule{0.283pt}{0.400pt}}
\multiput(600.59,338.65)(0.482,-0.581){9}{\rule{0.116pt}{0.567pt}}
\multiput(599.17,339.82)(6.000,-5.824){2}{\rule{0.400pt}{0.283pt}}
\multiput(606.00,332.93)(0.491,-0.482){9}{\rule{0.500pt}{0.116pt}}
\multiput(606.00,333.17)(4.962,-6.000){2}{\rule{0.250pt}{0.400pt}}
\multiput(612.59,325.65)(0.482,-0.581){9}{\rule{0.116pt}{0.567pt}}
\multiput(611.17,326.82)(6.000,-5.824){2}{\rule{0.400pt}{0.283pt}}
\multiput(618.00,319.93)(0.491,-0.482){9}{\rule{0.500pt}{0.116pt}}
\multiput(618.00,320.17)(4.962,-6.000){2}{\rule{0.250pt}{0.400pt}}
\multiput(624.00,313.93)(0.581,-0.482){9}{\rule{0.567pt}{0.116pt}}
\multiput(624.00,314.17)(5.824,-6.000){2}{\rule{0.283pt}{0.400pt}}
\multiput(631.00,307.93)(0.491,-0.482){9}{\rule{0.500pt}{0.116pt}}
\multiput(631.00,308.17)(4.962,-6.000){2}{\rule{0.250pt}{0.400pt}}
\multiput(637.00,301.93)(0.491,-0.482){9}{\rule{0.500pt}{0.116pt}}
\multiput(637.00,302.17)(4.962,-6.000){2}{\rule{0.250pt}{0.400pt}}
\multiput(643.00,295.93)(0.491,-0.482){9}{\rule{0.500pt}{0.116pt}}
\multiput(643.00,296.17)(4.962,-6.000){2}{\rule{0.250pt}{0.400pt}}
\multiput(649.00,289.93)(0.581,-0.482){9}{\rule{0.567pt}{0.116pt}}
\multiput(649.00,290.17)(5.824,-6.000){2}{\rule{0.283pt}{0.400pt}}
\multiput(656.00,283.93)(0.491,-0.482){9}{\rule{0.500pt}{0.116pt}}
\multiput(656.00,284.17)(4.962,-6.000){2}{\rule{0.250pt}{0.400pt}}
\multiput(662.00,277.93)(0.491,-0.482){9}{\rule{0.500pt}{0.116pt}}
\multiput(662.00,278.17)(4.962,-6.000){2}{\rule{0.250pt}{0.400pt}}
\multiput(668.00,271.93)(0.491,-0.482){9}{\rule{0.500pt}{0.116pt}}
\multiput(668.00,272.17)(4.962,-6.000){2}{\rule{0.250pt}{0.400pt}}
\multiput(674.00,265.93)(0.491,-0.482){9}{\rule{0.500pt}{0.116pt}}
\multiput(674.00,266.17)(4.962,-6.000){2}{\rule{0.250pt}{0.400pt}}
\multiput(680.00,259.93)(0.710,-0.477){7}{\rule{0.660pt}{0.115pt}}
\multiput(680.00,260.17)(5.630,-5.000){2}{\rule{0.330pt}{0.400pt}}
\multiput(687.00,254.93)(0.491,-0.482){9}{\rule{0.500pt}{0.116pt}}
\multiput(687.00,255.17)(4.962,-6.000){2}{\rule{0.250pt}{0.400pt}}
\multiput(693.00,248.93)(0.491,-0.482){9}{\rule{0.500pt}{0.116pt}}
\multiput(693.00,249.17)(4.962,-6.000){2}{\rule{0.250pt}{0.400pt}}
\multiput(699.00,242.93)(0.599,-0.477){7}{\rule{0.580pt}{0.115pt}}
\multiput(699.00,243.17)(4.796,-5.000){2}{\rule{0.290pt}{0.400pt}}
\multiput(705.00,237.93)(0.581,-0.482){9}{\rule{0.567pt}{0.116pt}}
\multiput(705.00,238.17)(5.824,-6.000){2}{\rule{0.283pt}{0.400pt}}
\multiput(712.00,231.93)(0.599,-0.477){7}{\rule{0.580pt}{0.115pt}}
\multiput(712.00,232.17)(4.796,-5.000){2}{\rule{0.290pt}{0.400pt}}
\multiput(718.00,226.93)(0.599,-0.477){7}{\rule{0.580pt}{0.115pt}}
\multiput(718.00,227.17)(4.796,-5.000){2}{\rule{0.290pt}{0.400pt}}
\multiput(724.00,221.93)(0.491,-0.482){9}{\rule{0.500pt}{0.116pt}}
\multiput(724.00,222.17)(4.962,-6.000){2}{\rule{0.250pt}{0.400pt}}
\multiput(730.00,215.93)(0.599,-0.477){7}{\rule{0.580pt}{0.115pt}}
\multiput(730.00,216.17)(4.796,-5.000){2}{\rule{0.290pt}{0.400pt}}
\multiput(736.00,210.93)(0.710,-0.477){7}{\rule{0.660pt}{0.115pt}}
\multiput(736.00,211.17)(5.630,-5.000){2}{\rule{0.330pt}{0.400pt}}
\multiput(743.00,205.93)(0.599,-0.477){7}{\rule{0.580pt}{0.115pt}}
\multiput(743.00,206.17)(4.796,-5.000){2}{\rule{0.290pt}{0.400pt}}
\multiput(749.00,200.93)(0.491,-0.482){9}{\rule{0.500pt}{0.116pt}}
\multiput(749.00,201.17)(4.962,-6.000){2}{\rule{0.250pt}{0.400pt}}
\multiput(755.00,194.93)(0.599,-0.477){7}{\rule{0.580pt}{0.115pt}}
\multiput(755.00,195.17)(4.796,-5.000){2}{\rule{0.290pt}{0.400pt}}
\multiput(761.00,189.93)(0.710,-0.477){7}{\rule{0.660pt}{0.115pt}}
\multiput(761.00,190.17)(5.630,-5.000){2}{\rule{0.330pt}{0.400pt}}
\multiput(768.00,184.93)(0.599,-0.477){7}{\rule{0.580pt}{0.115pt}}
\multiput(768.00,185.17)(4.796,-5.000){2}{\rule{0.290pt}{0.400pt}}
\multiput(774.00,179.93)(0.599,-0.477){7}{\rule{0.580pt}{0.115pt}}
\multiput(774.00,180.17)(4.796,-5.000){2}{\rule{0.290pt}{0.400pt}}
\multiput(780.00,174.93)(0.599,-0.477){7}{\rule{0.580pt}{0.115pt}}
\multiput(780.00,175.17)(4.796,-5.000){2}{\rule{0.290pt}{0.400pt}}
\multiput(786.00,169.93)(0.599,-0.477){7}{\rule{0.580pt}{0.115pt}}
\multiput(786.00,170.17)(4.796,-5.000){2}{\rule{0.290pt}{0.400pt}}
\multiput(792.00,164.94)(0.920,-0.468){5}{\rule{0.800pt}{0.113pt}}
\multiput(792.00,165.17)(5.340,-4.000){2}{\rule{0.400pt}{0.400pt}}
\multiput(799.00,160.93)(0.599,-0.477){7}{\rule{0.580pt}{0.115pt}}
\multiput(799.00,161.17)(4.796,-5.000){2}{\rule{0.290pt}{0.400pt}}
\multiput(805.00,155.93)(0.599,-0.477){7}{\rule{0.580pt}{0.115pt}}
\multiput(805.00,156.17)(4.796,-5.000){2}{\rule{0.290pt}{0.400pt}}
\multiput(811.00,150.93)(0.599,-0.477){7}{\rule{0.580pt}{0.115pt}}
\multiput(811.00,151.17)(4.796,-5.000){2}{\rule{0.290pt}{0.400pt}}
\multiput(817.00,145.93)(0.710,-0.477){7}{\rule{0.660pt}{0.115pt}}
\multiput(817.00,146.17)(5.630,-5.000){2}{\rule{0.330pt}{0.400pt}}
\multiput(824.00,140.94)(0.774,-0.468){5}{\rule{0.700pt}{0.113pt}}
\multiput(824.00,141.17)(4.547,-4.000){2}{\rule{0.350pt}{0.400pt}}
\multiput(830.00,136.93)(0.599,-0.477){7}{\rule{0.580pt}{0.115pt}}
\multiput(830.00,137.17)(4.796,-5.000){2}{\rule{0.290pt}{0.400pt}}
\put(299,804){\usebox{\plotpoint}}
\put(299.0,774.0){\rule[-0.200pt]{0.400pt}{7.227pt}}
\put(336,730){\usebox{\plotpoint}}
\put(336.0,699.0){\rule[-0.200pt]{0.400pt}{7.468pt}}
\put(385,651){\usebox{\plotpoint}}
\put(385.0,611.0){\rule[-0.200pt]{0.400pt}{9.636pt}}
\put(434,547){\usebox{\plotpoint}}
\put(434.0,502.0){\rule[-0.200pt]{0.400pt}{10.840pt}}
\put(484,529){\usebox{\plotpoint}}
\put(484.0,467.0){\rule[-0.200pt]{0.400pt}{14.936pt}}
\put(540,429){\usebox{\plotpoint}}
\put(540.0,368.0){\rule[-0.200pt]{0.400pt}{14.695pt}}
\put(587,405){\usebox{\plotpoint}}
\put(587.0,326.0){\rule[-0.200pt]{0.400pt}{19.031pt}}
\put(663,287){\usebox{\plotpoint}}
\put(663.0,212.0){\rule[-0.200pt]{0.400pt}{18.067pt}}
\put(761,205){\usebox{\plotpoint}}
\put(761.0,118.0){\rule[-0.200pt]{0.400pt}{20.958pt}}
\put(299,791){\raisebox{-.8pt}{\makebox(0,0){$\Box$}}}
\put(336,712){\raisebox{-.8pt}{\makebox(0,0){$\Box$}}}
\put(385,632){\raisebox{-.8pt}{\makebox(0,0){$\Box$}}}
\put(434,524){\raisebox{-.8pt}{\makebox(0,0){$\Box$}}}
\put(484,497){\raisebox{-.8pt}{\makebox(0,0){$\Box$}}}
\put(540,399){\raisebox{-.8pt}{\makebox(0,0){$\Box$}}}
\put(587,375){\raisebox{-.8pt}{\makebox(0,0){$\Box$}}}
\put(663,256){\raisebox{-.8pt}{\makebox(0,0){$\Box$}}}
\put(761,170){\raisebox{-.8pt}{\makebox(0,0){$\Box$}}}
\end{picture}